\documentclass[a4paper,11pt]{article}
\usepackage{nfraconf,graphicx,cite,amssymb}

\begin{document}
\baselineskip=13pt

\title{WEAK GRAVITATIONAL LENSING WITH SKA}

\author{Peter Schneider}
\address{Max-Planck-Institut f. Astrophysik\\
P.O. Box 1523\\
D-85740 Garching\\
Germany\\
E-mail: peter@mpa-garching.mpg.de}


\maketitle
\def\s{{({\rm s})}}
\def\eps{\epsilon}
\def\ts{\thinspace}
\def\d{{\rm d}}
\def\ave#1{\left\langle #1 \right\rangle}
\def\vec#1{{\bf #1}}

\abstract{I shall outline the basic principles and some observational
aspects of weak gravitational lensing, and discuss several
applications of this powerful tool in observational cosmology. It will
be explained why the applications have been restricted to optical
observations up to now, and why SKA is going to change this. I
conclude with a few general remarks on a comparison between SKA and
the NGST, both being facilities which will provide a tremendous step
forward in radio and near-IR astronomy, respectively, into completely
unknown territory.}

\section{Introduction}
Gravitational light deflection provides astronomers with a unique tool
to study the mass distribution in the Universe. The resulting
phenomena, called gravitational lensing, have attracted great interest
in recent years, since gravitational light deflection is independent
of the nature and the state of the matter which causes the deflection.
Hence in particular, lensing is sensitive to the dark matter, and is
therefore extensively employed for investigating the dark matter
distribution, on scales ranging from stellar-mass objects (as done in
microlensing studies), galaxy-mass scales (as in multiply-imaged
AGNs), to clusters (e.g., through giant luminous arcs) and
larger-scale matter distribution. Lensing is popular also because the
underlying physics is very simple (it is just gravity, hence only
General Relativity is needed, and in most cases, the correct answers
are obtained from quasi-Newtonian theory, with all deflection angles
multiplied by a factor 2), in marked contrast to nearly all other
methods used for understanding the mass distribution in the Universe.

The phenomena mentioned above can be summarized as ``strong
gravitational lensing''. For them, the lensing phenomenon is readily
detected in individual sources, e.g., the characteristic brightness
variation of one background star signifies the occurrence of
microlensing, or the strongly distorted images (arcs) of background
galaxies seen in a foreground galaxy cluster are sufficient to
identify the lensing property of the deflector. Strong lensing
therefore yields strong constraints on the deflecting mass
distribution: for example, the mass of a lensing galaxy must be such
that the observed multiple images are mapped back to the same source
position; in particular, the total projected mass of the lensing
galaxy in a circle tracing the multiple images is very well
constrained then.

If a galaxy cluster is strong enough to distort the images of one or a
few background galaxies very strongly, it appears obvious that for
larger radial distances from the cluster center, the images of
background galaxies are distorted as well, though weaker. If all
galaxies were circular, this smaller distortion could be measured on
individual images as well. However, galaxies have an intrinsic
ellipticity which prevents the measurement of weak image distortions
from individual images. But if one considers an ensemble of
background galaxies whose intrinsic orientation is random, the weak
distortion will impose a coherent alignment of the images which is
measurable, provided the distortion is larger than $\sim
\sigma_\eps/\sqrt{N}$, where $\sigma_\eps$ is the intrinsic rms
ellipticity of the background galaxies, and $N$ is the number of
images from which the shear shall be measured. The accuracy with which
the average distortion over a small solid angle can be measured
therefore depends on the number density of sources for which the shape
measurement can be carried out. Whereas the optical sky is densely
covered with faint galaxies (at the magnitude limit at which these
{\it weak lensing} measurements are typically carried out, the number
density of galaxies is $\sim 30/{\rm arcmin}^2$), the radio sky is
literally empty (cf. the number of optical and radio sources in the
HDF)! Therefore, weak lensing has up to now been restricted to optical
astronomy (with the exception of the work by Kamionkowski et
al. 1999). The SKA will change this situation drastically: the
expected number density of radio sources routinely achieved with SKA
is expected to be at least as large as that currently obtainable with
deep ground-based optical observations.

In this contribution, I shall describe the basic theory of (weak)
lensing in Sect.\ts 2, and discuss a few technicalities in Sect.\ts
3. Then, some applications of weak lensing are discussed in Sect.\ts
4. I shall then try to put SKA into perspective with respect to weak
lensing, and close with a few general remarks.

\section{Theory of (weak) lensing}
The basic theory of gravitational lensing can be found in our monograph
\cite{SEF}, or in several more recent review articles
(e.g., \cite{FM,BN}). Here I shall
just state a few general relations needed for the following
discussion.

As light is deflected by a single mass distribution at distance $D_{\rm
d}$, a source at distance $D_{\rm s}$ which, without an intervening
mass distribution would be seen at an angular position $\vec\beta$,
will be seen at a slightly different position $\vec\theta$, such that
the lens equation $\vec\beta=\vec\theta-\vec\alpha(\vec\theta)$ is
satisfied. Here, $\vec\alpha(\vec\theta)$ is the (scaled) deflection
angle which depends on the impact vector $\vec\theta$ of the light ray,
and on the mass distribution. Drawing the analogy to Newtonian
gravity, $\vec\alpha$ can be considered as a `force field'. Indeed, it
can be obtained as the gradient of a potential, the so-called
deflection potential $\psi(\vec\theta)$. The source of this potential
is the surface mass density of the deflector $\Sigma(\vec\theta)$, or
its dimensionless analogon,
$$
\kappa(\vec\theta)={4\pi D_{\rm d} D_{\rm ds}\over c^2 D_{\rm
s}}\Sigma(\vec\theta)\; .
\eqno (1)
$$
Here, $D_{\rm ds}$ is the distance from the lens to the source. All
these distances are to be interpreted as angular-diameter distances,
and they depend, for given source and lens redshifts, on the
cosmological model.  The deflection potential obeys a Poisson-like
equation, $\nabla^2\psi=2\kappa$. Multiple images of a source occur if
the lens equation has more than one solution $\vec\theta$ for a given
source position $\vec\beta$.

Light is not only deflected as a whole, but undergoes
differential deflection which distorts the size and shape of
light bundles. If the size of the source is much smaller than the
characteristic angular scale of the deflector, one can locally
linearize the lens mapping, which is then described by the Jacobian of
the lens equation
$$
{\cal A}(\vec\theta)=\pmatrix{1-\kappa-\gamma_1 & -\gamma_2 \cr
-\gamma_2 & 1-\kappa +\gamma_1 \cr }
=(1-\kappa)\pmatrix{1-g_1 & -g_2 \cr -g_2 & 1+g_1 \cr }\; ,
\eqno (2)
$$
Here, $\gamma_i$ are the two components of the tidal gravitational
field, called {\it shear},
which is described by the trace-free part of the Jacobian (as in
Newtonian theory), and which are conveniently written as a complex
number,
$\gamma=\gamma_1+{\rm i}\gamma_2=(\psi_{,11}-\psi_{,22})/2 +{\rm
i}\psi_{,12}$, and indices separated by a comma denote partial
derivatives. The final version of ${\cal A}$ in (2) shows that the
shape distortion of images is described solely by the {\it reduced
shear} $g:=\gamma/(1-\kappa)$. The change of the apparent flux of an
image of a small source is given by the local area distortion of the
lens mapping, 
$\mu=[(1-\kappa)^2-|\gamma|^2]^{-1}$, where the magnification $\mu$ is
the ratio of the observed flux of an image and that of the unlensed
source. Giant arcs are obtained at places where one of the eigenvalues
of ${\cal A}$ is close to zero, i.e., where $\mu\gg 1$.

In the next section we shall describe how one can obtain the (reduced)
shear from observed image ellipticities. Suppose we have measured
$\gamma$ over some field (say, around a cluster). Since $\gamma$ and
$\psi$ are linearly related, one can determine $\psi$ from $\gamma$,
and from that, the surface mass density can be obtained. Hence, the
measurement of the shear can be used directly to obtain the underlying
surface mass distribution.

\section{Technicalities}
The theory of weak lensing is well understood, as is the application
of this effect to several astrophysical situations, as will be
described later. The main difficulty of applying these concepts to
observational data lies in the measurement of what can be called
``ellipticity''. Since the SKA may be well superiour to optical
observations in this regard, some of the technical issues shall be
described here. 

\subsection{The principle of shear measurements}
The isophotes of faint sources are not elliptical in general, and
hence, one must define a quantity which characterizes the shape of an
object and how this shape quantity is affected by shear. A convenient
way to proceed, at least for theoreticians, 
is to define the tensor of second brightness moments,
$$
Q_{ij}=\int\d^2\theta\;I(\vec\theta)\,
(\theta_i-\bar\theta_i)\,(\theta_j -\bar\theta_j) \;,
\eqno (3)
$$
where $I(\vec\theta)$ is the surface brightness of an image, and
$\bar{\vec\theta}$ is the center of the image, defined such that the dipole
vector [which is defined in analogy to (3)] vanishes. Similarly, one
can define the tensor $Q^\s_{ij}$ of second brightness moments
of the corresponding source. The lens equation then relates these two
tensors by
$$
Q^\s=\det{\cal A}\,({\cal A}\,Q\,{\cal A})\; .
\eqno (4)
$$
The (complex) ellipticity $\eps$ is now defined in terms of $Q$, as
$$
\eps={Q_{11}-Q_{22} + 2{\rm i} Q_{12} \over
Q_{11}+Q_{22}+2\sqrt{Q_{11} Q_{22}-Q_{12}^2}}\quad,
\eqno (5)
$$
and correspondingly the ellipticity $\eps^\s$ of the intrinsic
brightness profile of the galaxy in terms of $Q_{ij}^\s$.
For example,
if an image has elliptical contours of axis ratio $r\le 1$, then
$|\eps|=(1-r)/(1+r)$.
From (4) one then derives the transformation between
intrinsic and observed ellipticity
$$
\eps^\s={\eps -g\over 1-g^* \eps}\;,
\eqno (6)
$$
where the asterisk denotes complex conjugation.
Finally, since the sources are assumed to be randomly oriented
intrinsically (and this assumption is not seriously challenged -- note
that the faint galaxies used for weak lensing come from a large range
of redshifts, and so most of them are not at all physically related),
the expectation value of $\eps^\s$ vanishes; this can be used to show
that the expectation value of the image ellipticity is just the
reduced shear $g$. Hence, the ellipticity of every galaxy image yields an
unbiased (though very noisy) estimate of the reduced shear. So much
for theory ...

\subsection{Practical measurement of the shear}
Unfortunately, the scheme presented above cannot be applied in real
situations, for several reasons: 
\begin{itemize}
\item Observations are noisy; e.g., CCD
images have a (near Poisson) noise due to the finite number of photons
per pixel. The brightness profile of an object falls off, and outside
its characteristic size, $I(\vec\theta)$ becomes dominated by noise. The
large weighting given by (3) to large $|\vec\theta-{\bar{\vec\theta}}|$
then means that the integral is completely dominated by noise. In
order to prevent this, one needs to limit the range of
integration. This can be achieved in several ways: either by adding a
weight function into the integrand which depends on the surface
brightness $I$, or by a weight function depending explicitly on
$|\vec\theta-{\bar{\vec\theta}}|$. The former choice renders the
relation between $I$ and $Q$ non-linear, and so the effects of noise
are more complicated to account for. Therefore, in the scheme
described below, a weight function $W(|\vec\theta-{\bar{\vec\theta}}|)$
is chosen. In addition, since galaxies are not isolated, one needs a
cut-off in the integration before the neighboring object enters the
integration range. The size of the weight function should be chosen
such as to match the size of the image. With a spatially dependent
weight function in the definition of $Q$, the transformation between
source and image plane is no longer given by the simple relation (4)
or (6).

\item Atmospheric turbulence
causes seeing; i.e., the point-spread function (PSF) for ground-based
optical astronomy is limited by the atmosphere. The PSF is further
affected by the imaging properties of the telescope. In any case,
since the size of faint galaxies is comparable to the size of the
seeing disk, images are severely smeared, i.e., they appear rounder
than they would be without seeing. In effect, seeing degrades the
measured shear. 

\item Owing to imperfect imaging properties in the
telescope, tracking errors, wind shake, or other reasons, the PSF is
not necessarily circularly symmetric. Even a circular image would then
appear elongated if observed through an anisotropic PSF. Hence, an
anisotropic PSF mimics shear. 
\end{itemize}

These effects have to be accounted for in order to obtain a reliable
estimate of the shear. Several image analysis tools have been written
specifically for weak lensing applications. We shall now indicate the
result of the necessary correction steps, as used in the {\sl
imcat} method \cite{KSB,LK,HFKS}. 
The observed image ellipticity $\eps^{(\rm obs)}$
(defined as in (4) and (6), but with a spatial weight function added)
can be written as
$$
\eps^{(\rm obs)}=\eps^0 + P^{\rm sm} q +P^{\rm g} g\;,
\eqno (7)
$$
where $\eps^0$ is the image ellipticity one would observe if there
were no shear, and if the PSF were isotropic. Although the relation
between $\eps^0$ and the intrinsic source ellipticity is complicated,
this is irrelevant in this scheme: if the expectation value of
$\eps^\s$ is zero, then necessarily that of $\eps^0$ also
vanishes. For small values of the PSF anisotropy and the shear, the
difference between $\eps^{(\rm obs)}$ and $\eps^0$ can be linearized,
which has been done in (7): $g$ is the reduced shear, and $q$ is the
ellipticity of the PSF. The two factors (they are indeed tensors in
general, but in practice turn out to be close to diagonal) $P^{\rm
sm}$ and $P^{\rm g}$ describe the response of the image ellipticity to
a PSF anisotropy and to a shear, respectively. These response factors
depend on the brightness profile of the images and have to be
calculated for each individual image. For example, a big image
responds less to a PSF anisotropy than a small image, and for a small
image, the response to a shear will be smaller than for a big one,
owing to the smearing by the PSF. Appropriately averaging (7) over a
set of galaxy images, using that the expectation value of $\eps^0$
vanishes, yields an estimate of the reduced shear.

The PSF anisotropy $q$ can be obtained by considering stellar
images. Light from a star is unaffected by shear, and in addition, if
smeared by an isotropic PSF, its image is round. Hence, for a star,
$\eps_*^0=0$ and $P_*^{\rm g}=0$, yielding $q=\eps_*^{(\rm
obs)}/P_*^{\rm sm}$. Provided that the number of stars in the data field is
sufficiently large, and that $q$ is a slowly varying function of
position on the CCD image, one can use the $q$'s as measured from the
stars to fit a (low-order) polynomial which then yields $q$ for
every position on the frame. This determination of $q$ works quite
well for some imaging instruments.

In addition, optical images show features which
have to be dealt with carefully if shear measurements are to be
carried out: cosmic rays have to be removed, diffraction spikes,
trails and bleeding can lead to artificial object detections which
characteristically have large moduli of the ellipticity etc. Whereas
these artefacts can be recognized and thus corrected for or avoided,
they complicate the weak lensing analysis. 

\section{Applications of weak lensing}
Weak lensing is a young research area (the first arclets have been
identified by Fort et al. in 1988 \cite{F88}, and the first coherent
shear around clusters was published by Tyson et al. in 1990
\cite{TVW}), and the range of potential applications has grown at a
substantial rate in the past few years. I shall outline a few of those
in this section, without claiming that this list is even approximately
complete; see \cite{M99} for an excellent review of weak lensing.

\subsection{Mass and mass distribution of clusters of galaxies}
Clusters of galaxies are the most massive bound structures in the
Universe, and one might therefore expect that these systems produce
the strongest weak lensing effect. Indeed, the vast majority of
successful weak lensing observations have been targeted towards
clusters. The results of these studies are two-dimensional maps of the
projected mass density, and estimates of the bulk mass of the cluster.
As mentioned at the end of Sect.\ 2, if the shear can be measured, the
surface mass density $\kappa(\vec\theta)$ can be reconstructed, due to
the linear relations between $\kappa$, $\psi$ and $\gamma$. However,
since a uniform surface mass density does not give rise to a shear
contribution (as in Newtonian gravity!), $\kappa$ is determined only
up to an additive constant. In addition, we have seen that not the
shear itself is an observable, but only the reduced shear $g$ can be
observed. This then leaves the invariance transformation 
$\kappa(\vec\theta) \to \lambda \kappa(\vec\theta)+(1-\lambda)$
\cite{F85,SS95}, which
for $\lambda\approx 1$ is similar to the addition of a homogeneous
mass density. The scaling factor $\lambda$ cannot be determined from
shear observations only, and is usually set such that, at least for
large data fields, $\kappa\approx 0$ near the boundary of the field,
or far from the cluster center. The mass reconstruction from the
observable reduced shear is not more difficult than it would be from
the shear itself (e.g., \cite{SS95b, SS96}).

Another way to quantify mass properties of clusters consists in
defining the filtered mass density
$$
M_{\rm ap}(\vec\theta_0):=\int\d^2\theta\; \kappa(\vec\theta)
\,U(|\vec\theta-\vec\theta_0|) \;,
\eqno (8)
$$
where $U(\theta)$ is a compensated filter,
$\int\d\theta\,\theta\,U(\theta)=0$ (this condition makes $M_{\rm ap}$
independent of the addition of a homogenous mass sheet) which vanishes
beyond a filter radius $R$. One can then show \cite{K94,S96}
that the aperture mass can also be expressed as
$$
M_{\rm ap}(\vec\theta_0)=\int\d^2\theta\;\gamma_{\rm
t}(\vec\theta_0;\vec\theta) \,Q(|\vec\theta|)\;,
\eqno (9)
$$
where $\gamma_{\rm t}(\vec\theta_0;\vec\theta)$ is the component of the
shear at position $\vec\theta_0+\vec\theta$ in the direction tangent to
the direction of $\vec\theta$, and $Q$ is a filter function which can
be calculated directly in terms of $U$, and which also vanishes for
$|\vec\theta|>R$.  Putting aside for a moment the difference between
shear and reduced shear, the latter expression for $M_{\rm ap}$ can be
estimated directly from observations by replacing the integral over
the shear by a sum over the (tangential component) of image
ellipticities. One example for the choice of $U$ is the so-called
$\zeta$-statistics \cite{K95}, where $U$ is a positive constant for
$0\le \theta\le \theta_1$, and a negative constant for $\theta_1\le
\theta\le R$, chosen such as to satisfy the compensation condition;
then, $Q$ is different from zero only in the annulus $\theta_1\le
\theta\le R$, and if $\theta_1$ is chosen sufficiently large as to
avoid the central part of the cluster where $\kappa$ takes appreciable
values, the difference between $g$ and $\gamma$ in that annulus is
small. For that choice, $M_{\rm ap}$ becomes the mean surface mass
density within the circle $\theta_1$, minus the mean density in the
annulus. Since the latter cannot be negative, this method yields a
lower limit of the cluster mass within $\theta_1$.

Cluster mass profiles have been reconstructed for about 20 clusters so
far (see \cite{M99} for a recent review), with data and methods
which span quite in range in quality. The resolution of these mass
maps depends on the number density of background galaxies used for the
shear measurement, and is highest for observations taken with the
HST. Several of the most detailed mass maps indicate that the bright
cluster galaxies follow the underlying mass distribution quite well
(e.g., \cite{SKSS, HFKS, K99}).
The detection of strong weak lensing signals in high-redshift
clusters ($z\sim 0.8$, see \cite{LK, CL98})
shows that they are genuine massive structures which may pose very
strong constraints on the cosmological parameters (e.g., \cite{BA}).

The above mentioned invariance transformation leaves the reduced shear
invariant, but the magnification changes according to $\mu \to
\mu/\lambda^2$, so that by measuring a magnification the invariance
can be broken. Observable magnification effects include the change of
the local number counts of faint galaxies (magnification bias) and the
change in image size at fixed surface brightness. The former of these
effects \cite{BTP}
has been observed in at least two clusters \cite{FMDF,T98}, whereas
the latter \cite{BN95} may be
difficult to observe from the ground (due to the PSF), but may be
detectable in space-based observations.

\subsection{Galaxy-galaxy lensing} 
Whereas clusters are massive enough to be studied individually with
weak lensing techniques, galaxies are not. However, if one is not
interested in the mass properties of an individual galaxy, but in the
statistical mass properties of an ensemble of galaxies, the weak
lensing signals of these galaxies can be statistically combined. If
one considers two populations of galaxies, foreground and background
galaxies, one would expect the ellipticities of the latter to be
aligned preferentially in the direction tangent to the nearest
foreground galaxies. Thus if one considers pairs of
foreground-background galaxies, one should see a signature of weak
lensing in the alignment statistics.

This idea can be formulated quantitatively, and the first
galaxy-galaxy lensing signal has been detected in \cite{BBS}, using a
single ground-based image. The foreground-background separation is
made statistically, on the basis of apparent magnitude -- fainter
galaxies are on average further away -- and the galaxy population is
parametrized according to Tully-Fischer-type relations. The analysis
found the characteristic rotational velocity of an $L_*$ galaxy to be
about 220\ts km/s, in accordance to what is known from their rotation
curves, and yielded a lower bound on the truncation radius of the dark
halo; here, $L_*$ is the characteristic luminosity scale entering the
Schechter luminosity function.
Although this latter result is not very powerful yet, future
observations using much larger data fields will be able to improve on
this dramatically. Further galaxy-galaxy lensing results have come
from HST observations (e.g., \cite{GR96,HU98}). With large enough
samples, the request for a parametrization of the galaxy population
may vanish, since then the galaxies can be binned according to
`similar types'. In this regard, the knowledge of even approximate
redshifts will be very useful, and photometric redshift techniques
seem to provide the necessary redshift accuracy.

\subsection{Detection of (dark) halos}
A weak lensing observation around a massive cluster leads to the
detection of a coherent shear signal. This signal is then used to
determine the mass of the cluster, in the way described above. As a
first step, though, one can ask whether the cluster is detected at all
in the weak lensing map, and this question can be quantified well (is
the tangential shear several $\sigma$ above the noise expected from
randomly oriented galaxy images?). Of course, one can also set up the
experiment in a slightly different way: by taking a wide-field image
in an arbitrary direction, one can look for points in that image
around which the tangential alignment of galaxy images is several
$\sigma$ above the noise level expected for randomly oriented
images. If that is the case (say in a 6-$\sigma$ situation), one would
conclude that this point corresponds to the `center' of a massive
halo. If at the same point an overdensity of galaxies were detected,
one would conclude to have detected a cluster -- selected by its mass
properties. If, on the other hand, this overdensity of galaxies is not
seen, then what? Would one conclude that a massive halo has been
detected which does not contain luminous galaxies, a cluster failed to
produce light? 

Before turning to this question, it should be pointed out that a
mass-selected sample of clusters would be of great value for
cosmology. Whereas usually clusters (like other astronomical objects)
are selected by the (optical or X-ray) light they emit, the comparison
with cosmological predictions (either semi-analytical or numerical) is
hampered by the fact that these concern the mass properties of objects
(halos) rather than the light. In order to make predictions of the
luminous properties, additional assumptions have to be made, such as a
mass-temperature relation for clusters, and its evolution with
redshift. The uncertainties associated with that have hampered the use
of the cluster abundance as a function of mass and redshift as a most
sensitive tool for testing cosmological models.\footnote{An additional
uncertainty of the association of X-ray selected clusters with
prediction from, e.g., Press-Schechter theory is that many of the
high-redshift clusters, which yield the most sensitive probe of the
cosmological model, do show very substantial substructure which
might indicate that the cluster formation is not yet completed for
them -- are those objects then to be identified with collapsed halos
as counted by Press-Schechter?} The selection of
clusters by weak lensing techniques (that is, by mass) would
circumvent these difficulties, and the observational results could be
compared directly with numerical N-body simulations of the evolution
of the dark matter in the Universe. 

A quantitative way to proceed is to use the aperture mass, defined in
(8) and (9) above. Positive values of $M_{\rm ap}$ signify the presence of a
mass peak, and the noise of $M_{\rm ap}$ can be determined either from
the data themselves (e.g., by repeated randomizations of the
orientation of images), or be estimated analytically. As an
illustrative example, consider a singular isothermal sphere with
velocity dispersion $\sigma_v$. The signal-to-noise ratio for its
detection is
$$
{{\rm S}\over {\rm N}}= 12.7 \left( {n\over 30{\rm arcmin}^{-2}}
 \right)^{1/2}
\left( {\sigma_\epsilon \over 0.2 }\right) ^{-1} 
\left(
 {\sigma_v\over 600{\rm km/s} }\right) ^{2}
\left( {\ln(\theta_2/\theta_1) \over \ln(10)}\right) ^{1/2}
\left\langle {D_{\rm ds}\over D_{\rm s}} \right\rangle \;
\eqno (10)
$$
where the angular bracket denotes an average over the source
population (the ratio $D_{\rm ds}/ D_{\rm s}$ is set to zero for
galaxies with smaller redshift than the lens), $\sigma_\eps$ is the
dispersion of the intrinsic ellipticity distribution, and $\theta_1$
and $\theta_2$ denote the inner and outer radius of an annulus in
which the shear is measured. This S/N is obtained if the filter
function $Q(\theta)$ is optimized for the mass profile of a singular
isothermal sphere; using more generic weight function, the S/N is reduced
by factors of order 1.5. Nevertheless, the foregoing estimate shows
that dark matter halos at intermediate redshift with velocity
dispersion in excess of $\sim 600{\rm km/s}$ can be detected from
their shear field, a prediction \cite{ME91, S96}
that was impressively verified in the case of the cluster MS1512
\cite{SSetal}. The expected abundance of such halos which
can be detected with that technique with ${\rm S/N}\ge 5$ depends on
the cosmology and the normalization of the power spectrum, but is in
excess of $\sim 10$ per square degree \cite{KS99}. This
abundance, and the corresponding $M_{\rm ap}$-spectrum, can be
directly compared with predictions from N-body simulations
\cite{RKJS}. Thus, even a modest wide-field survey of 
$\sim 10$ square degrees will lead to an extremely useful sample of
mass-selected halos.

Probably, most of them will be luminous, so that one detects ordinary
clusters. Then the redshift of these halos can be determined
spectroscopically, and one could investigate the range of
mass-to-light ratios for these clusters. If this range extends toward
very high M/L values then it may appear plausible that there may be
some halos for which this value even exceeds those for which it can be
measured, because there is so little light, and one therefore might
detect `dark' clusters (against which the normal selection procedure
clearly would bias).

A first example of a shear-detected matter concentration has been
found recently \cite{ERB}. With two independent data sets of deep
wide-field optical imaging, a tangential alignment was found around a
point which, by random orientation of the galaxy images, would be as
unlikely as $10^{-6}$. No obvious concentration of bright galaxies
centered on the mass peak is found; however, faint X-ray emission from
near the matter concentration has been identified from archival X-ray
data. The interpretation of this result is unclear at present, and
follow-up observations are needed to test whether the mass
concentration is associated with a high-redshift cluster, or is an
example of a `dark clump'.

\subsection{Cosmic shear and cosmology}
If, as considered in the previous section, a blank field is targeted
and investigated with respect to weak lensing, the strong tangential
alignments show up as the most prominent features which can be
identified with mass peaks. But of course, they cover only a small
fraction of the total area. In the rest of the field, the image
ellipticities are also affected by the tidal gravitational field
between us and the source galaxies and therefore carry information
about the larger-scale mass distribution in the Universe. Indeed, the
two-point statistics of the shear (such as 2-pt. correlation function,
or rms shear within circular apertures) is related directly to the
projected power spectrum of the mass distribution in the
Universe. Higher-order statistics correspondingly yields higher-order
statistical properties of the Large Scale Structure; e.g., the
skewness of the shear appears to be a sensitive probe of the density
parameter $\Omega$ \cite{BvWM, SvWJK, vWBM}. It should be noted that
the weak lensing investigation of the LSS is the only method currently
known which does not depend on assumptions about how the luminous
matter traces the underlying dark matter distribution, with the
exception of Cosmic Microwave Background experiments which study the
LSS at much higher redshifts ($z\sim 1000$) and at large comoving
length scale ($\gtrsim 10 {\rm Mpc}$).  Comparison between CMB and
weak lensing results can be turned into a sensitive test of the
gravitational instability picture of structure growth. In addition,
even modestly large weak lensing surveys can break the degeneracies of
cosmological model parameters left with the next generation of CMB
experiments \cite{HT}.

It must be mentioned, however, that the so-called cosmic shear is
weak; depending on cosmology, the rms shear on a scale of a few
arcminutes is about 1\%. This implies that the systematic effects
mentioned before (e.g., PSF anisotropy) must be understood to levels
well below 1\% to make a quantitative measurement of cosmic
shear. Whereas no sure-stopper has been identified up to now for
ground-based optical imaging, this field may especially profit from
the control one may expect over the PSF of SKA, in particular on large
angular scales.

\section{Weak lensing and SKA}
Let us summarize the observational requirements for an efficient weak
lensing study: 
\begin{itemize}
\item The number density $n$ of objects for which a shape
can be measured reliably should be as high as possible; 

\item the mean
redshift $\ave{z}$ of this source population should be high, to put a
large fraction of them into the background of the lenses to be
investigated and to maximize the average of $D_{\rm ds}/D_{\rm s}$;

\item the source population should be as round as possible, i.e.,
$\sigma_\eps$ should be minimized; 

\item the ratio of the size of the
PSF and that of the source should be as small as possible, to minimize
the corrections that have to be applied to the observed ellipticities;

\item for the same reason, the PSF anisotropy should be small; 

\item the PSF must be controllable, to allow the corrections for it; 

\item for
most of the applications listed in Sect.\ 4, the field-of-view should
be large to enhance the statistical significance of the results.
\end{itemize}

SKA will push radio astronomy in a position where the radio sky is as
much filled with sources as in current optical astronomy. Since the
limiting flux of SKA will be $\sim 100$ times fainter than currently
achievable, predictions about $n$ and $\ave{z}$ are quite
uncertain. If the dominant source population correspond to normal or
star-forming galaxies, then $\ave{z}\sim 1$. If an additional source
population turns up, potentially increasing $n$, their characteristic
redshift will be most important. The value of $\sigma_\eps$ of the
faint radio sources is also unknown; if they are dominated by core-jet
type of sources, then $\sigma_\eps$ can be quite high, whereas
hydrogen emission from normal galaxies yield probably much rounder
sources. The FOV of SKA is large, and its PSF will be well
controllable, perhaps better than it can ever be hoped for optical
imaging. Therefore, SKA may be able to measure shears smaller than is
possible with optical images.

Optical astronomy is also developing quickly, and to put SKA in
perspective, one has to consider the expected evolution of optical
imaging instruments in the next decade. In ground-based astronomy, two
developments are of particular interest for weak lensing: the
installment of 8-meter class telescopes, and the coming-on-line of
wide-field imaging cameras, both at excellent observing sites (see
\cite{AR98}). These developments will allow to tremendously improve
the weak lensing capabilities of optical astronomy, in particular in
wide-field applications. However, seeing provides a fundamental
limitation: as galaxies tend to become smaller when they are fainter,
the number density of objects for which a shape can be measured
reliably (i.e., corrected for seeing, which is only possible if it is
not much smaller than the seeing disk) is limited, to about
60/arcmin$^2$, as shown by deep Keck images \cite{CL98}. Space-based
observations can achieve much higher number densities, as shown in the
HDF, and the Advanced Camera on-board HST will yield substantial
progress to the field. But it, as well as imaging instruments
currently under discussion for the Next generation Space Telescope,
will have a fairly limited FOV, of order $4'\times 4'$, and at least
for wide-field application cannot easily compete with ground-based
wide-field imaging cameras. In this respect, SKA will play a very
important role for weak lensing, as it combines a high number density
of objects for which ellipticities can be measured reliably (due to
the knowledge of its PSF), with a large FOV -- again, provided
$\sigma_\eps$ is not excessively large and $\ave{z}$ not small for
these faint radio sources.

\section{General remarks}
SKA will be superiour to current radio observatories by orders of
magnitude, in several respects, in particular in sensitivity and
FOV. Coupled with the angular resolution, it will allow observations
one cannot even nearly approach today: Specifying the properties of
SKA in order to carry out specific observations is, at least in part,
wild guesswork. This specification is partly based on a
fainter-N'-further philosophy, i.e., one can study objects of the same
class at much lower luminosity (like faint AGNs) or similar objects at
larger distance (e.g., SNRs in more distant galaxies). Although these
may be valuable science drivers, this projection is very conservative.
I think there can be no doubt that right in the first
weeks of operation of SKA many new discoveries will be made (though
the meaning of them may become clear only much later, e.g., after
optical/IR spectroscopy and/or X-ray observations), since so much more
volume of the observational phase space will open up. Also, even in
the time span between now and the finishing of SKA, new science
drivers will be identified. 

Let me compare SKA with NGST which will also yield a jump in
sensitivity by orders of magnitude relative to current observations,
at least in the near-IR. Deepest K-band images today reach 24th
magnitude, whereas NGST routinely will go to $\sim$29th
magnitude. Predictions into that regime are of course very uncertain
extrapolations, but the specification of instrumental capabilities
needs to be taken from here. If one considers a possible distribution
of observations for NGST, as obtained from the current version of the
Design Reference Mission, then a large fraction goes into science
programs which would not have been proposed only 5 years ago: for
example, weak-lensing related imaging (5 years ago, the very first
weak lensing results were published), the detailed investigation of
the galaxy distribution by photometric redshifts (the U-dropout
technique provided the first sizeable sample some 3 years ago), or the
search for extra-solar planetary systems (with the first ones
discovered only a few years ago). No question that the DRM will change
in the course of time and may look very different at the time of
launch of NGST. One can even argue that NGST will provide answers to
question we have not yet dared to ask. One example would be the mass
distribution in clusters: each massive cluster will contain tens of
observable strong lensing features (arcs, multiple images) and a very
cleanly outlined weak lensing structure, so that the projected mass
distribution in these clusters can be obtained with very high angular
resolution. What do we actually learn from these mass maps? Which
cluster properties, or properties of cluster galaxies, can be probed,
which models for the mass distribution in clusters can be critically
checked (and eventually rejected) with these data?

The lesson to learn from that is that the current predictions of what one
wants to do with NGST and SKA {\it are very conservative indeed}. This
is to some degree unavoidable and can only partly be overcome with
model calculations and simulations. What is important, though, is to
construct these observatories such that they are flexible enough to
allow a major modification of scientific goals and the corresponding
observational programs. Since one does not know which direction these
future requirements will take, the best guess may be to cover as much
observational phase space as politically wise, financially
affortable and compatible with the key science issues, to have a
versatile instrument at hand. This aspect may be particularly relevant
for SKA which is likely to render most other radio telescopes
working in the same wavelength regime obsolete.

\section*{Acknowledgement}
I would like to thank the organizers for the kind invitation to this
meeting which I found very stimulating, and for their financial
support. This work was supported in part by the
"Sonderforschungsbereich 375-95 f\"ur Astro-Teilchenphysik" der
Deutschen Forschungsgemeinschaft.

\end{document}